\documentclass[11pt,oneside,letterpaper]{article}
\usepackage{amssymb}
\usepackage{amsmath}
\usepackage[dvips]{graphicx}
\usepackage{setspace}
\usepackage{amsfonts}
\usepackage{fancyhdr}
\usepackage{xcolor}
\usepackage{caption}
\usepackage{graphicx}
\usepackage{rotating}
\usepackage{comment}
\usepackage{color}
\usepackage{cite}
\usepackage{braket}
\definecolor{darkgreen}{rgb}{0,0.5,0}
\definecolor{darkblue}{rgb}{0,0,0.6}
\definecolor{purple}{rgb}{0.4,.2,0.7}

\newcommand{\f}{\frac}

\newcommand{\be}{\begin{equation}}
\newcommand{\ee}{\end{equation}}

\usepackage[colorlinks=true,citecolor=darkgreen,linkcolor=black,urlcolor=purple]{hyperref}

\usepackage{pdfsync}

\makeatletter
\newcommand*{\defeq}{\mathrel{\rlap{%
                     \raisebox{0.3ex}{$\m@th\cdot$}}%
                     \raisebox{-0.3ex}{$\m@th\cdot$}}%
                     =} 
\makeatother

\DeclareMathOperator{\Tr}{Tr}
\def\be{\begin{eqnarray}}
\def\ee{\end{eqnarray}}

\newcommand{\bea}{\begin{eqnarray}}
\newcommand{\eea}{\end{eqnarray}}
\def\ben{\begin{equation}}
\def\een{\end{equation}}

 \let\b=\beta   
   
 \let\m=\mu    \let\r=v
 \let\t=\tau

\let\f=\frac

\def\be{\begin{equation}}
\def\ee{\end{equation}}
\def\ba{\begin{array}}
\def\ea{\end{array}}

\def\ba#1\ea{\begin{align}#1\end{align}}
\def\bs#1\es{\begin{split}#1\end{split}}

\allowdisplaybreaks  

\begin{document}
\onehalfspacing

\begin{center}

~
\vskip5mm

{\LARGE  {
A symmetry principle for emergent spacetime
}}

\vskip10mm

Edgar Shaghoulian

\vskip5mm

{Department of Physics, Cornell University, Ithaca, New York, USA} 
\vskip5mm

\vskip5mm

\tt{ eshaghoulian@cornell.edu\\}


\end{center}

\vspace{4mm}

\begin{abstract}
\noindent
There are many examples where geometry and gravity are concepts that emerge from a theory of quantum mechanics without gravity. This suggests thinking of  gravity as an exotic phase of matter. Quantifying this phase in the Landau paradigm requires some sort of symmetry principle or order parameter that captures its appearance. In this essay we propose higher-form symmetries as a symmetry principle underlying emergent spacetime. We explore higher-form symmetries in gauge-gravity duality and explain how their breaking describes features of a gravitational theory. Such symmetries imply the existence of nonlocal objects in the gravitational theory -- in gauge-gravity duality  these are the strings and branes of the bulk theory -- giving an alternative way to understand the nonlocality necessary in any ultraviolet completion of gravity. \\
\\
\\
{\centering
\emph{Essay written for the Gravity Research Foundation 2020 Awards for Essays on Gravitation.}}
 \end{abstract}

\pagebreak
\pagestyle{plain}

\setcounter{tocdepth}{2}
{}
\vfill

\section*{Introduction}
We propose that weakly coupled gravity is an exotic phase of matter describable -- like any other phase of matter in the Landau paradigm -- by appropriate symmetries. There are several inescapable features of gravity this framework must reproduce. In this essay we focus on two particularly important ones. The first is the geometric nature of black hole entropy. The second is the triviality of correlation functions on quotient spacetimes. These hallmarks of weakly coupled gravity have straightforward derivations and conceptual underpinnings in such a theory. How are these inherently geometric features supposed to be encoded in a non-geometric theory of quantum mechanics?

In gauge-gravity duality \cite{Itzhaki:1998dd}, the weakly coupled limit is given by the large-$N$ limit of the quantum mechanics, where $N$ is some measure of the degrees of freedom.  We propose that the existence of certain higher-form symmetries and their symmetry-breaking pattern encodes the features above at large $N$. This principle does not assume conformal symmetry or supersymmetry, yet provides a dynamical mechanism for the sparse spectrum assumed by the conformal bootstrap \cite{Heemskerk:2009pn}. Furthermore, higher-form symmetries imply nonlocal objects in the gravitational description, suggesting that such objects are necessary for an ultraviolet completion of gravity. 

\section*{Black hole entropy}
The entropy of a black hole in Einstein gravity is given by $S = \f{A}{4G}$ \cite{Bekenstein:1973ur, Hawking:1974sw}. The entropy in a  general theory of gravity with higher curvature corrections is similarly given by the integral of a local quantity over the event horizon \cite{Wald:1993nt}. This means that under a quotient of the horizon $\mathcal{M}^{d-1}$ by some group $\Gamma$, the entropy changes by the appropriate factor:
\be
\mathcal{M}^{d-1}\rightarrow \mathcal{M}^{d-1}/\Gamma \qquad \implies \qquad S \rightarrow S/|\Gamma|\,.
\ee
In particular the entropy density is invariant. This feature is also responsible for the remarkable thermal phase structure of gravity \cite{Hartman:2014oaa, Belin:2016yll}.

\section*{Correlation functions in gravity}
The propagator of any weakly coupled field is a Green's function, meaning it can be found on a quotient spacetime by the method of images. The atoms of a general Feynman diagram are these propagators, which in the semiclassical limit implies an equality between the connected correlation function in the quotient spacetime and the original spacetime \cite{Shaghoulian:2016xbx}. As an example, the connected two-point function on spacetime $\mathcal{M}^{d-1}\times \mathbb{R}$ gives the connected two-point function on $\mathcal{M}^{d-1}\times S^1_L$:
\be
\langle O(\vec{y},x) O(0)\rangle_{S^1_L} = \sum_{n=-\infty}^{\infty} \langle O(\vec{y}, x+nL) O(0)\rangle_{\mathbb{R}}\,,\qquad x \in S^1_L\,.
\ee
$S^1_L$ is a circle of circumference $L$. Thus the finite-size corrections in the semiclassical limit are trivial. The analogous statement is true for connected $n$-point correlators in arbitrary quotient spacetimes. 

\section*{Non-gravitational origin}
These features, which are rather straightforward to derive in the geometric description, need to be encoded in the quantum-mechanical theory from which they emerge. We now argue that they are reproduced by the existence of a higher-form symmetry and a particular pattern of symmetry breaking. 

A higher-form (i.e. $p$-form with $p>1$) symmetry is like an ordinary ($p=1$) symmetry, except that the operators charged under the symmetry are nonlocal, with spatial dimension $p-1$.
We focus on one-form $\mathbb{Z}_N$ center symmetry in large-$N$ $SU(N)$ gauge theory. Consider the background manifold to have a non-contractible cycle, say $\mathcal{M}^{d-1}\times S^1_\b$. The symmetry transformation acts on the gauge field $A_\m$ as
\be
A_\m \rightarrow g (A_\m + \partial_\m)g^{-1}\,.
\ee
Unlike a gauge transformation, the group element $g: \mathcal{M}^{d-1} \times S^1_\b \rightarrow SU(N)$ has a twist along the circle, $g(\t + \b) = g(\t)h$ for $h \in \mathbb{Z}_N$. While local fields are invariant under this transformation, a Wilson line wrapping the circle transforms as 
\be
W(C) = \Tr P \exp\left[\int_C  A_\m dx^\m\right] \longrightarrow h W(C)\,.
\ee
Thus, a nonzero expectation value $\langle W(C)\rangle$ implies a spontaneous breaking of center symmetry along $C$.

\begin{figure}
\begin{center}
\includegraphics[scale=.54]{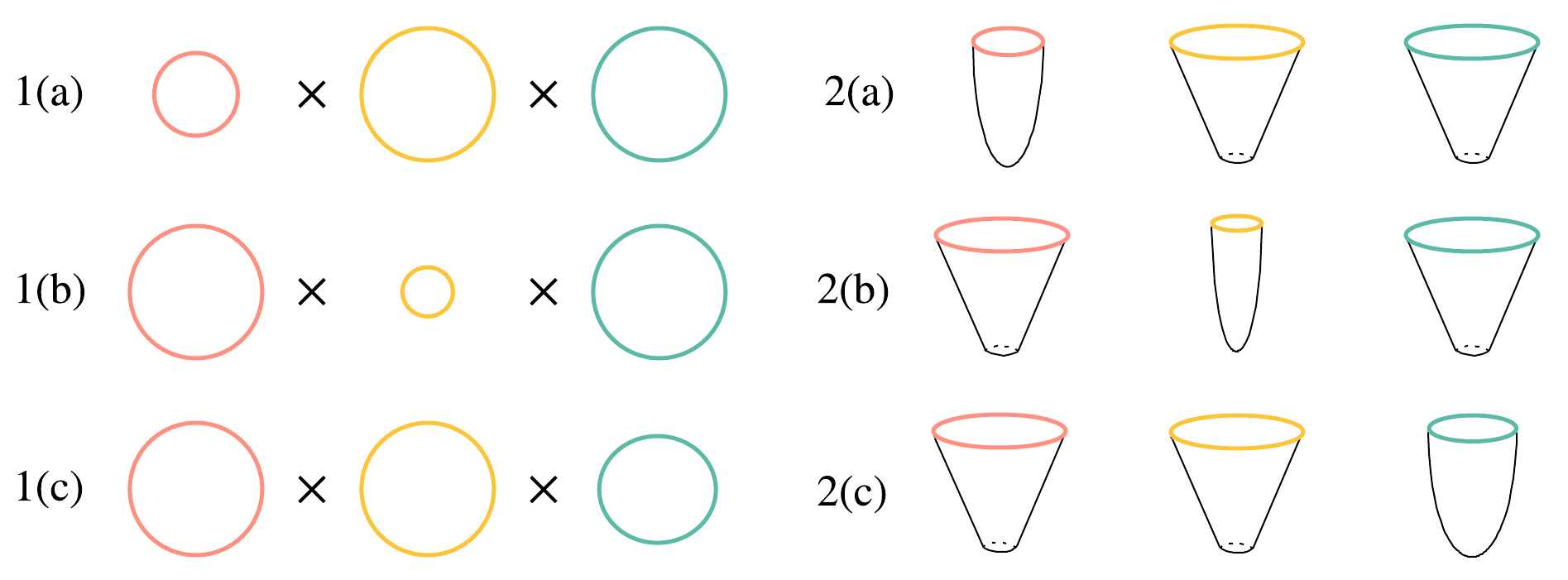}
\end{center}
\captionsetup{labelformat=empty}
\caption{Figure: 1(a)-(c) represent three different configurations of the spacetime manifold $S^1 \times S^1 \times S^1$ of a three-dimensional QFT. This theory has an emergent four-dimensional gravitational description where one of the three cycles is contractible in the bulk, represented in 2(a)-(c) (the additional dimension in the gravitational description is redundantly displayed for each cycle; the topology of the gravitational spacetime is $S^1 \times S^1 \times D$ where $D$ is a disk). As the cycle lengths are varied, thermal and quantum phase transitions occur taking us between 2(a)-(c). $\langle W(C)\rangle = 0$ if and only if the cycle being wrapped is non-contractible in the bulk.}
\end{figure}

In gauge-gravity duality, the maximally supersymmetric version of this theory with $d \leq 6$ has an emergent gravitational description. For $d=4$ this is just $\mathcal{N}=4$ super Yang-Mills, which has an emergent gravitational description on AdS$_5 \times S^5$. For $d\neq 4$ the theory is non-conformal but has the one-form symmetry discussed above. The geometry of the field theory $\mathcal{M}^d$ is encoded in the timelike boundary of the gravitational spacetime $\mathcal{M}^{d+1}$. The cycle $C$ can be contractible in the bulk, in which case gauge-gravity duality tells us $\langle W(C)\rangle \neq 0$ (and $\langle W(C) \rangle = 0$ otherwise) \cite{Maldacena:1998im}, see figure above.

There is an important observation about such symmetries, which is a theorem in many contexts \cite{Eguchi:1982nm, Kovtun:2004bz, Shaghoulian:2016xbx}:

\emph{If translation and center symmetry are preserved along a non-contractible cycle in $\mathcal{M}^d$, then appropriate observables are independent of the size of the cycle at leading order in $N$.}

The intuition for such a statement comes from the physics of confinement. In a large-$N$ confined theory, center symmetry is unbroken along the thermal circle $S^1_\b$, and the confined phase degrees of freedom are glueballs and potentially mesons. The interactions of these composites are suppressed by $1/N$, so at leading order in $N$ they cannot interact with their images to learn of their finite-size universe. 

This remarkable effect precisely reproduces the geometric nature of gravity discussed above. This is because cycles coming from smooth gravitational quotients are non-contractible, implying $\langle W(C) \rangle = 0$ and preservation of center symmetry. The observables which are independent of cycle size include the entropy density and certain frequency-space correlators. The former is precisely what we observed with black hole entropy, while the latter implies position-space correlators in quotient spacetimes are given by a sum over images in the covering space.

 For a smooth, translation-invariant gravitational description, we know that only one cycle is contractible in the bulk. This is the particular symmetry-breaking pattern implied by gravity. The only known field theories with such a symmetry-breaking pattern have an emergent gravitational description. Exactly which cycle in the field theory is contractible in the bulk depends on various parameters, see figure above.

In gauge-gravity duality, the one-form center symmetry implies one-dimensional extended objects in the bulk. These are precisely the fundamental strings of the bulk string theory. The relevant symmetry can be of even higher form: in the example of AdS$_7$/CFT$_6$ coming from a stack of M$5$-branes, the CFT has a two-form symmetry which implies the existence of two-dimensional objects in the bulk, the M$2$-branes. In many of these examples, changing the pattern of higher-form symmetry symmetry breaking leads to the loss of a geometric description. We turn to this next. 

\section*{Two parables: introducing geometry through center symmetry}
An illuminating example is the duality between the $O(N)$ vector model and a nonlocal, higher-spin theory in AdS$_4$. The vector model has no higher-form symmetries, and the higher-spin theory does not have the geometric features discussed above. The authors of \cite{Chang:2012kt} manipulated the higher-spin theory  to bind together the higher-spin ``bits" into the strings of ABJ theory \cite{Aharony:2008gk}, which has a local, geometric limit. From the point of view of the vector model, this construction has an incredibly simple interpretation: center symmetry is introduced!

The particular \emph{pattern} of higher-form symmetry breaking is also indispensable. For example, at weak coupling in maximally supersymmetric Yang-Mills, there is a higher-form symmetry but there is no emergent local geometry. The required symmetry-breaking pattern  -- and an emergent geometry to boot -- only appear at strong coupling \cite{Myers:1999psa}.

\section*{The unreasonable effectiveness of Euclidean gravity}
This framework also sheds light on the unreasonable effectiveness of Euclidean gravity. Hawking explained the entropy of gravitational spacetimes in Euclidean gravity as follows: for spacetimes with a contractible thermal circle, the on-shell action $-\log Z(\b)$ depends nontrivially on $\b$ and implies a thermal entropy, $S(\beta) = (1-\beta \partial_\beta)\log Z(\beta)$. On the other hand, if the thermal circle is non-contractible, the on-shell action is proportional to $\beta$ and $S(\b) = 0$. This simple topological analysis is precisely what captures the preservation or breaking of a one-form symmetry and provides a quantum-mechanical origin to Hawking's geometric argument. When the symmetry is broken, the free energy density depends on $\b$ and the thermal entropy is nonzero. Alternatively, when the symmetry is preserved, the free energy density is independent of $\b$ and the thermal entropy therefore vanishes.

A recent development further illustrating the power of Euclidean gravity is a derivation of the unitarity of Hawking radiation through replica wormholes \cite{Almheiri:2019qdq, Penington:2019kki}. These works imply that the interior of the black hole is redundantly encoded in the Hawking radiation located far away. This points to a huge nonlocality of the information flow in a radiating black hole, even though the starting point was a local, weakly-coupled theory of gravity. This seemingly paradoxical situation mirrors what we confronted above, where we invoked nonlocal objects behind the scenes of a local, emergent geometry. Perhaps the same nonlocal ingredients are behind the scenes of unitary evolution in a black hole background.

\section*{Discussion}
In this essay we have proposed that gravity is an exotic phase of matter, identifiable by a particular pattern of higher-form symmetry breaking. Further details on this perspective can be found in \cite{Shaghoulian:2016xbx}. Beyond the technical puzzles elaborated in the text, this perspective sheds light on some conceptual puzzles of gauge-gravity duality, including the necessity of bulk strings/branes (answer: they are implied by higher-form symmetries) and the prevalence of \emph{gauge} symmetry in the quantum-mechanical origin of spacetime (answer: gauge theories often come equipped with higher-form symmetries). 

\section*{Acknowledgments}
I would like to thank Daniel Harlow for encouraging me to write this essay and Tom Hartman for suggestions on it.  I would also like to thank Aleksey Cherman, David Gross, Nabil Iqbal, Joseph Polchinski, and Leonard Susskind for conversations related to these ideas.
I am supported by the Simons Foundation as part of the Simons Collaboration on the Nonperturbative Bootstrap.

\small
\bibliographystyle{ourbst}
\bibliography{HigherFormBib}

\end{document}